\documentclass[prb,twocolumn,showpacs,citeautoscript,balancelastpage]{revtex4-1}

\usepackage{amsmath,amssymb,graphicx}
\usepackage{color}
\usepackage{xspace}
\usepackage{bm}
\usepackage[extra]{tipa}
\usepackage{dcolumn}

\setcitestyle{square,numbers}

\newcommand{\las}[0]{\langle}
\newcommand{\ras}[0]{\rangle}
\newcommand{\llas}[0]{\langle\langle}
\newcommand{\rras}[0]{\rangle\rangle}
\newcommand{\la}[0]{\left\las}
\newcommand{\ra}[0]{\right\ras}

\newcommand{\ie}[0]{i.e.\@\xspace}

\newcommand{\rmi}{\text{i}}
\newcommand{\rmd}{\text{d}}
\newcommand{\UP}[0]{\uparrow}
\newcommand{\DO}[0]{\downarrow}

\newcommand{\om}[0]{\omega}

\newcommand{\kF}{k_\text{F}}

\newcommand{\nag}{{\phantom{\dag}}}

\begin{document}

\title{Rashba coupling and magnetic order in correlated helical liquids}

\author{Martin Hohenadler} \author{Fakher F. Assaad}

\affiliation{\mbox{Institut f\"ur Theoretische Physik und Astrophysik,
    Universit\"at W\"urzburg, Am Hubland, 97074 W\"urzburg, Germany}}

\begin{abstract}
  We study strongly correlated helical liquids with and without Rashba
  coupling using quantum Monte Carlo simulations of the Kane-Mele model with
  a Hubbard interaction at the edge. Independent of the Rashba coupling, we
  find that interactions enhance spin correlations and suppress the
  spectral weight at the Fermi level. For sufficiently strong interactions, a gap can be observed in the
  single-particle spectral function. However, based on a finite-size
  scaling analysis and theoretical arguments, we argue that this gap is
  closed by order parameter fluctuations in the Luttinger liquid phase even
  at zero temperature, and filled in by thermally induced kinks in the order
  parameter in the Mott phase at finite temperatures. While the
  bosonization suggests an umklapp-driven Mott transition only in the
  presence of Rashba coupling and hence a significant impact of the latter,
  our numerical results are almost unaffected by Rashba coupling even at low temperatures.
\end{abstract}

\date{\today}



\maketitle

\section{Introduction}\label{sec:introduction}

Helical edge states are a hallmark feature of quantum spin Hall insulators
\cite{KaMe05,HaKa10}. In the absence of electronic correlations,
there is a direct (bulk-boundary) correspondence between the number of
helical edge state pairs and the $Z_2$ topological invariant
\cite{KaMe05}. Each pair forms a Kramers doublet related by time-reversal
symmetry, with a symmetry-protected crossing at one of the two time-reversal
invariant points in the Brillouin zone. A single pair of edge states is
therefore stable with respect to disorder and weak interactions. The (nearly)
quantized spin Hall conductivity can be used to detect the topological state
in experiments \cite{BeHuZh06,Koenig07,PhysRevLett.95.136602}.

The interplay of a topological band structure and electronic correlations has
emerged as a very fruitful and active field of research. For a sufficiently
strong electron-electron repulsion, quantum spin Hall insulators can undergo
a Mott transition to a topologically trivial antiferromagnet without edge
states \cite{RaHu10}. Quantum Monte Carlo simulations
\cite{Hohenadler10,Ho.Me.La.We.Mu.As.12,Zh.Wu.Zh.11} of the Kane-Mele-Hubbard
model have confirmed the expected 3D XY universality class of this
transition, which remains unchanged even for a long-range Coulomb interaction
\cite{PhysRevB.90.085146}. Correlation effects have been studied in great
detail also for other models~\cite{HoAsreview2013}.

Because the bulk of a quantum spin Hall insulator is gapped, the edge states
are exponentially localized at the edge. At sufficiently low energies, they
may therefore be considered to be one-dimensional. Consequently, the effect
of electron-electron interactions cannot be captured using perturbation
theory or mean-field methods. Instead, a faithful description of correlated
helical edge states is based on the bosonization method
\cite{Voit94,Giamarchi}.

A remarkable theoretical prediction is the occurrence of an edge-Mott
transition as a result of strong electron-electron repulsion
\cite{Wu06}. Such a transition requires half-filled edge states and a Rashba
spin-orbit term to allow umklapp scattering \cite{Ho.As.11}. Whereas the
first condition is rather special, the second is generic for experimental
realizations. Above a critical interaction (the critical point corresponds to
a Luttinger parameter $K=1/2$), um{\-}klapp scattering is a relevant
perturbation, and the edge is expected to exhibit Ising long-range magnetic
order and an excitation gap at $T=0$ \cite{Wu06}. In this scenario,
time-reversal symmetry is broken only at the edge while the bulk remains
topological, leading to a breakdown of the bulk-boundary correspondence
\cite{PhysRevB.74.195312,PhysRevB.74.045125,PhysRevB.83.085426}.

In contrast to the bulk Mott transition, the edge-Mott transition has not
been demonstrated numerically. In particular, the Rashba term gives rise to a
sign problem that makes quantum Monte Carlo simulations challenging.
Previous work on models without Rashba coupling (and hence no Mott
transition) revealed a strong suppression (enhancement) of the charge (spin)
fluctuations and of the spectral weight at the Fermi level suggestive of 
a Mott transition
\cite{Hohenadler10,Ho.As.11,Zh.Wu.Zh.11,PhysRevB.83.205122}. It was also
shown that correlation effects are generically stronger at the edges than in
the bulk \cite{Hohenadler10,PhysRevLett.107.166806}, and that the condition
$K<1/2$ for an edge Mott state seems to be fulfilled in large regions of the
phase diagram \cite{Hohenadler10,Ho.As.11,Zh.Wu.Zh.11}.  Correlated edge
states were also simulated numerically with quantum cluster methods
\cite{Yu.Xie.Li.11,Wu.Ra.Li.LH.11,Wa.Da.Xi.12,La.Re.Th.Ra.13}, but the latter
do not capture the characteristic Luttinger liquid physics. For
future experiments, it is interesting to consider how different the
signatures of the Mott transition are from generic interaction effects.

Here, we simulate a Kane-Mele-Hubbard model
using a quantum Carlo (QMC) method. Because interactions are only taken into
account at the edge, the sign problem caused by the Rashba term is not prohibitive.  By comparing spin
correlation functions and single-particle spectral functions we conclude that
for accessible system sizes
the differences due to Rashba coupling are much
more subtle than expected. The main reason for the absence of striking
signatures of this transition is that correlation effects strongly modify the
helical edge states even in the absence of a Mott transition.

The rest of this paper is organized as follows. In
Sec.~\ref{sec:modelandmethod}, we discuss the model and the method used. Our
results are presented in Sec.~\ref{sec:results}. A discussion of our findings
is given in Sec.~\ref{sec:discussion}, followed by our conclusions in
Sec.~\ref{sec:conclusions}.

\section{Model}\label{sec:modelandmethod}

The Kane-Mele-Hubbard model \cite{RaHu10} provides a framework to study
correlation effects in two-dimensional topological insulators (see
Ref.~\cite{HoAsreview2013} for a review). The Hamiltonian of the underlying
Kane-Mele model is \cite{KaMe05}
\begin{eqnarray}\label{eq:KM}
  H_{\mbox{\scriptsize{KM}}} 
  = 
  &-&t \sum_{\las i,j \ras} \hat{c}^{\dagger}_{i} \hat{c}^\nag_{j} 
  + 
  \rmi\,\lambda \sum_{\llas i,j\rras}
  \hat{c}^{\dagger}_{i}\,
  (\boldsymbol{\nu}_{ij} \cdot \boldsymbol{\sigma})\,
  \hat{c}^\nag_{j} 
  \\\nonumber
  &+&
  \rmi\,\alpha \sum_{\las i,j\ras} 
  \hat{c}^{\dagger}_{i}\,
  (\boldsymbol{\sigma} \times \hat{\bm{d}}^\nag_{ij})\cdot \hat{\bm{z}} \,\hat{c}^\nag_{j}\,,
\end{eqnarray}
where we have used the spinor notation $\hat{c}_i =
\left(c_{i\UP},c_{i\DO}\right)^T$ and the Pauli vector
$\bm{\sigma}$. Hamiltonian~(\ref{eq:KM}) consists of the usual tight-binding
hopping term familiar from graphene ($\sim t$) \cite{Neto_rev}, the
$s^z$-conserving spin-orbit term first derived by Kane and Mele ($\sim
\lambda$) \cite{KaMe05}, and a Rashba term ($\sim \alpha$)
\cite{Rasha}. All three terms correspond to single-electron hopping processes
on the honeycomb lattice. The sign of the spin-orbit and Rashba terms depends
on the sublattice, the direction of the hop (left or right turn), and the
spin. Explicitly, $\boldsymbol{\nu}_{ij} = \bm{d}_{ik} \times
\bm{d}_{kj}/|\bm{d}_{ik} \times \bm{d}_{kj}|$ and
$\hat{\bm{d}}_{ij}=\bm{d}_{ij}/|\bm{d}_{ij}|$, with $\bm{d}_{ij}$ being a
vector connecting sites $i$ and $j$, and $k$ the intermediate site on the
hopping path from $i$ to $j$.  For
$\alpha=0$, the Kane-Mele model~(\ref{eq:KM}) is equivalent to two
time-reversed copies of the Haldane model \cite{Haldane98}.

For $\lambda=\alpha=0$, Hamiltonian~(\ref{eq:KM}) describes a semimetal with
a linear spectrum at the Dirac points and a vanishing density of
states at the Fermi level \cite{Neto_rev}. If $\alpha=0$, the Kane-Mele model has a
quantum spin Hall ground state for any $\lambda>0$, characterized by a $Z_2$
topological invariant $\nu=1$, a quantized spin Hall conductivity
$\sigma^\text{s}_{xy}=\nu e^2/2\pi$, and a pair of helical edge
states. The Rashba term does not conserve spin [reducing the spin symmetry
from $U(1)$ to $Z_2$], and competes with the $\lambda$ term. The quantum
spin Hall state remains stable up to $\alpha=2\sqrt{3}\lambda$
\cite{PhysRevLett.97.036808,PhysRevLett.95.136602}.

The phase diagram of the Kane-Mele-Hubbard model with an interaction
$H_U=\frac{U}{2} \sum_{{i}} (\hat{c}^{\dagger}_{{i}} \hat{c}^\nag_{{i}} -
1)^2$, but without a Rashba term, is known from mean-field theory
\cite{RaHu10} and QMC simulations \cite{PhysRevB.90.085146,Zh.Wu.Zh.11}.  For
$\lambda>0$, the ground state is a quantum spin Hall insulator for
$U<U_\text{c}(\lambda)$, and an antiferromagnetic insulator with long-range
order in the transverse spin direction for $U\geq
U_\text{c}(\lambda)$. Importantly, the quantum spin Hall state at $U>0$ is
adiabatically connected to the $U=0$ state
\cite{Hohenadler10,Ho.Me.La.We.Mu.As.12}, and bulk interactions are of minor
importance in this gapped phase. While an additional Rashba term competes
with the topological band gap induced by the spin-orbit term, recent cluster
calculations suggest that the critical $U$ for the bulk Mott transition is
only weakly affected \cite{La.Re.Th.Ra.13}.
 
To compare with theoretical predictions, it is necessary to
consider sufficiently large ribbons at low temperatures. The ribbon width
$L_\perp$ (the extent perpendicular to the edge) has to be large enough to
avoid a hybridization of opposite edges
\cite{PhysRevLett.103.166403,PhysRevLett.107.166806,PhysRevB.85.165138}. The
length $L$ determines the momentum resolution and the maximal length
scale for correlations.  To compare to the bosonization results, we want to avoid any
approximations that break symmetries or neglect fluctuations. QMC simulations
of the full Kane-Mele-Hubbard model cannot take into account a Rashba term
due to a sign problem \cite{Ho.Me.La.We.Mu.As.12}. Therefore, we make use of
the previously introduced idea \cite{Hohenadler10,Ho.As.11} of considering
electronic interactions only at the edge, an approximation that is well
justified by the absence of low-energy bulk excitations and the exponential
localization of the edge states. This simplification allows us to simulate
correlated edge states with Rashba coupling on reasonably large lattices.  
The Rashba term gives rise to a sign problem. However, the latter is
not prohibitive here because it depends on the volume of the interacting
one-dimensional edge ($L$) rather than the volume of the two-dimensional
ribbon ($L\times L_\perp$).
We will show in Sec.~\ref{sec:results} that our simulations reproduce the
theoretical predictions, and hence capture the physics of correlated helical
liquids. Keeping interactions only at the edge excludes the bulk magnetic
transition. The latter is also not described by the strictly one-dimensional
bosonization, although it can be incorporated by an {\it ad hoc} mass term
proportional to the bulk magnetization \cite{Wu.Ra.Li.LH.11}.

The Kane-Mele-Hubbard edge model with Rashba coupling is defined by the
action
\begin{align}\nonumber \label{eq:action}
  S = &-\sum_{ij} \iint_{0}^{\beta} d\tau d\tau'\,
  \hat{c}^{\dagger}_{i}(\tau) {\bm{G}^{-1}_{0}} (i-j,\tau-\tau')
  \hat{c}^\nag_{j}(\tau')
  \\
  &\hspace*{4em}+ \frac{U}{2} \sum_{i} \int_{0}^{\beta} d\tau
  \left[\hat{c}^{\dagger}_{{i}}(\tau) \hat{c}^\nag_{{i}}(\tau) -1
  \right]^2\,,
\end{align}
where $i,j$ index edge sites \cite{Hohenadler10,Ho.As.11}, and $\bm{G}_{0}$
is the $2\times 2$ Nambu Green function matrix of the Kane-Mele
model~(\ref{eq:KM}) on an $L\times L_\perp$ zigzag
ribbon. Equation~(\ref{eq:action}) describes correlated helical edge states
of a topological insulator coupled to the two-dimensional, noninteracting
bulk. It can be solved exactly using the CT-INT QMC method \cite{Rubtsov05},
and the numerical effort is independent of the ribbon width $L_\perp$
\cite{Hohenadler10,Ho.As.11}. 

Here, we will use two related variants of the CT-INT method, namely the
original, finite-temperature method \cite{Rubtsov05} with inverse temperature
$\beta$, and a projective formulation \cite{Assaad07} with projection
parameter $\theta$.  Periodic (open) boundary conditions were used in the
direction parallel (perpendicular) to the edge.

\section{Numerical results}\label{sec:results}

This section is organized as follows. We first briefly consider the
noninteracting case as a reference point for our study of interaction
effects. We then study the evolution of the real-space spin and charge
correlations as a function of $U$, before turning to the single-particle
spectral function. We use $t$ as the energy unit, set the lattice constant
and $k_\text{B}$ to one, and consider a half-filled lattice.

\begin{figure}[t]
  \includegraphics[width=0.45\textwidth]{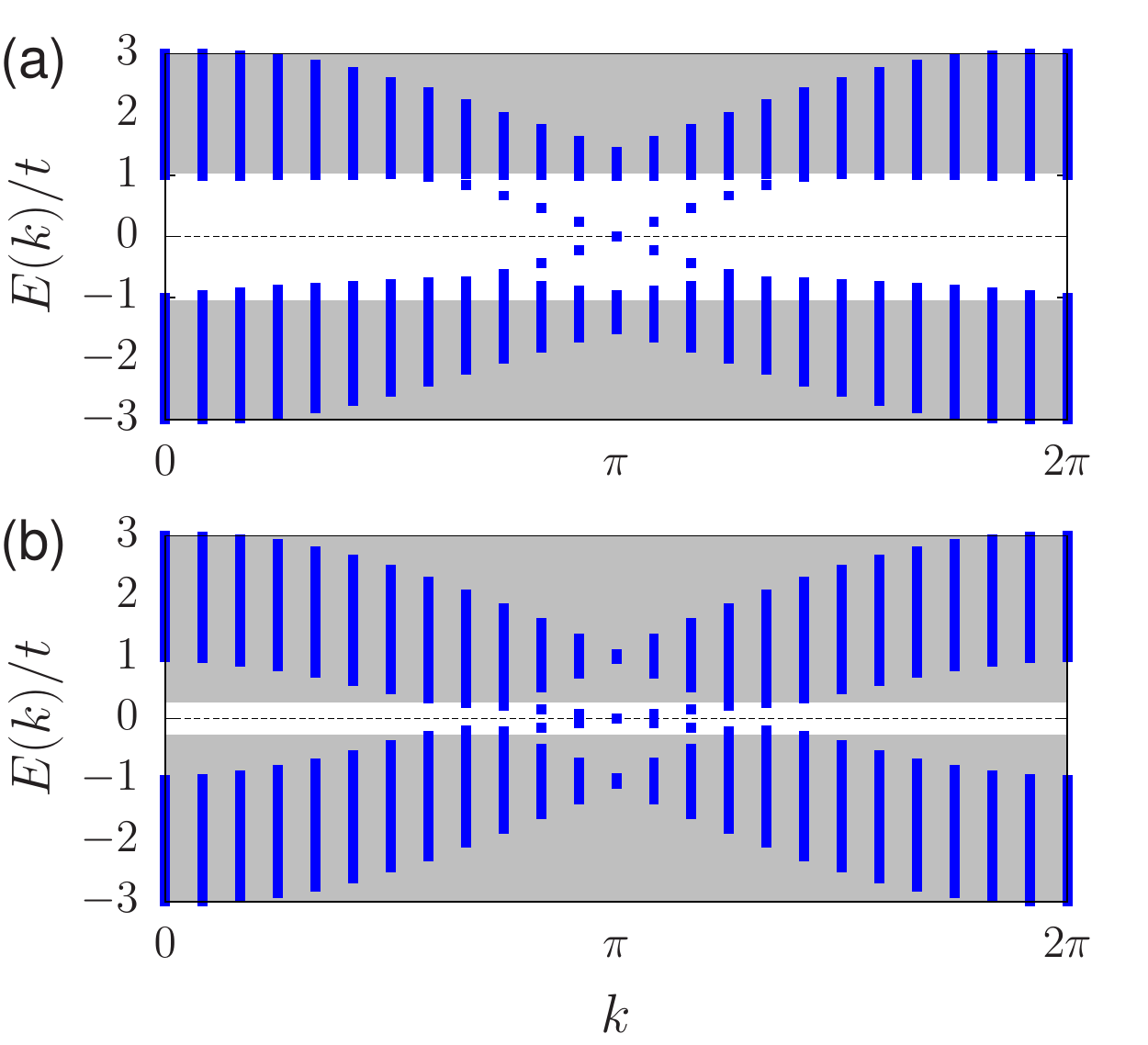}
  \caption{\label{fig:spectraU0} (Color online) Spectrum of the
    noninteracting Kane-Mele model on an $L\times L_\perp$ zigzag ribbon,
    showing helical edge states crossing at $k=\pi$. (a) $\lambda/t=0.2$, (b)
    $\lambda/t=0.05$. Here, $\alpha=\lambda$, $L=24$, and $L_\perp=64$. The
    shaded areas correspond to bulk states outside the $\alpha=0$ band gap
    $\Delta_\text{SO}=3\sqrt{3}\lambda$; the dashed lines indicate the Fermi
    energy.  }
\end{figure}

\subsection{Noninteracting case}

Although we are mostly interested in the regime of strong correlations, the
noninteracting case $U=0$ provides an important limit in which we can verify
that (for suitable parameters) our numerical results reproduce the analytical
predictions.

Figure~\ref{fig:spectraU0} shows the eigenvalue spectrum of a zigzag ribbon
with dimensions $L\times L_\perp = 24\times64$, for $\lambda/t=\alpha/t=0.2$.
The ribbon is wide enough to make inter-edge tunneling irrelevant
\cite{PhysRevLett.103.166403,PhysRevLett.107.166806,PhysRevB.85.165138}.
Figure~\ref{fig:spectraU0}(a) reveals the familiar spin-orbit gap and a pair
of helical edge states with a protected crossing at $k=\pi$
\cite{KaMe05}.  The gap is of the order of $t$ (the exact value for
$\alpha=0$ is $\Delta_\text{SO}=3\sqrt{3}\lambda\approx1.04t$
\cite{KaMe05}). The gray shading indicates the separation between low-energy
edge states and high-energy bulk states. The Rashba coupling reduces the band
gap \cite{KaMe05}, and eliminates the particle-hole symmetry. The choice of
$\lambda/t=0.2$ produces a large gap and hence minimizes the impact of the
bulk states in simulations, which is crucial to reach the low-energy limit
where the bosonization results are valid. In the latter, bulk states are
neglected, whereas they are fully taken into account in our simulations.
To make the impact of Rashba coupling visible we chose a substantial
value $\alpha=\lambda$ that preserves the quantum spin Hall state.

In addition to $\lambda/t=0.2$, we shall also present selected results for
the case $\lambda/t=0.05$, for which the spectrum is shown in
Fig.~\ref{fig:spectraU0}(b). The spin-orbit gap is much smaller
($\Delta_\text{SO}/t\approx0.26$), and the edge states have a significantly
smaller velocity. The smaller gap makes it harder to reach the low-energy
regime, but the smaller velocity enables us to simulate stronger
correlations. (The relevant parameters controlling correlations are $U$ and
the edge state velocity $v_\text{F}$ that scales with $\lambda$ \cite{Ho.As.11}.)

\begin{figure}[t]
  \includegraphics[width=0.45\textwidth]{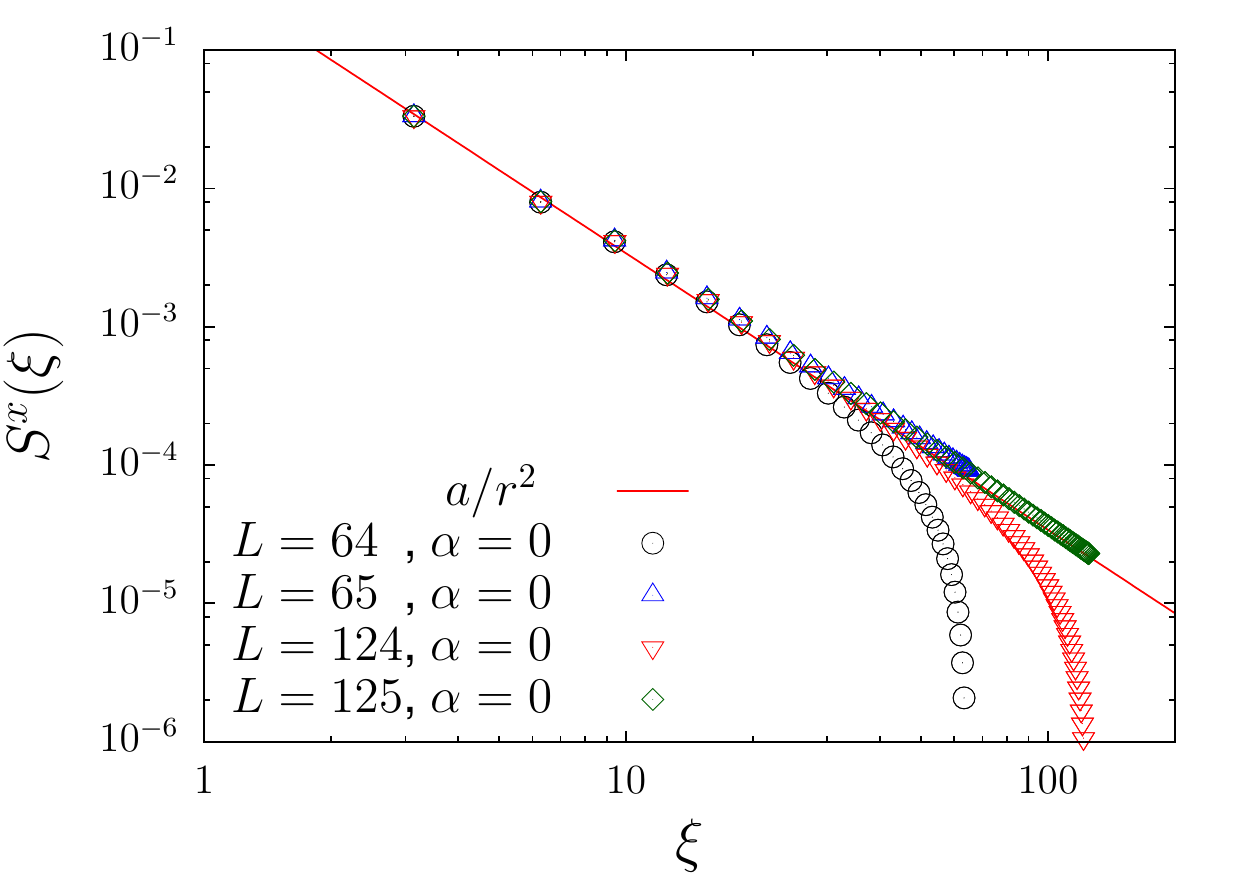}
  \caption{\label{fig:correlationsU0} (Color online) Transverse spin
    correlations for $U=0$ and different $L$. Here, $\lambda/t=0.2$, $\beta
    t=200$, and $L_\perp=64$. The solid line illustrates the expected $1/r^2$
    decay.}
\end{figure}

For $U=0$, the amplitude of transverse spin correlations is expected to decay
as $1/r^2$ at sufficiently large distances, as follows from the bosonization
result
\begin{equation}\label{eq:spincorrelator}
  S^{x}(r)= \las s^{x}(r) s^{x}(0) \ras 
  \sim
  \cos(2\kF r ) \,r^{-2K}
\end{equation}
by setting the Luttinger liquid interaction parameter $K$ equal to one.  We
reproduce this behavior in Fig.~\ref{fig:correlationsU0}. To remove the effects of the periodic boundary
conditions, we plot the conformal distance
\cite{Cardybook}
\begin{equation}
  \xi=L\sin(\pi r/L)\,.
\end{equation}

\begin{figure}[t]
  \includegraphics[width=0.45\textwidth]{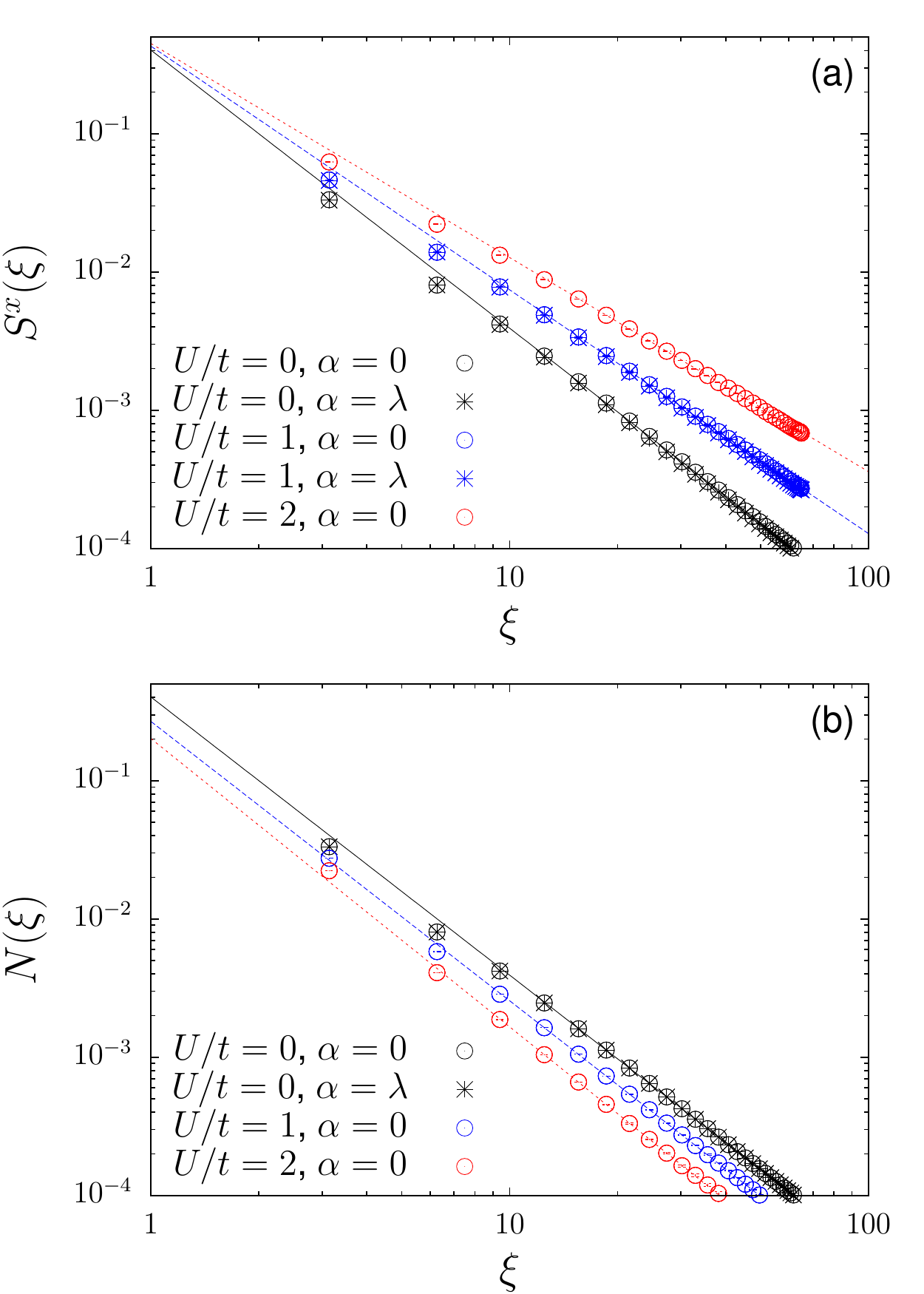}
  \caption{\label{fig:correlationsU} (Color online) Projective CT-INT results
    for the transverse spin and charge correlations for different
    interactions $U$. Here, $\lambda/t=0.2$, $L=65$, $L_\perp=64$, and $\theta t
    = 40$. Lines are fits to the form $a/\xi^\gamma$ at large distances.}
\end{figure}

\begin{figure}[t]
  \includegraphics[width=0.45\textwidth]{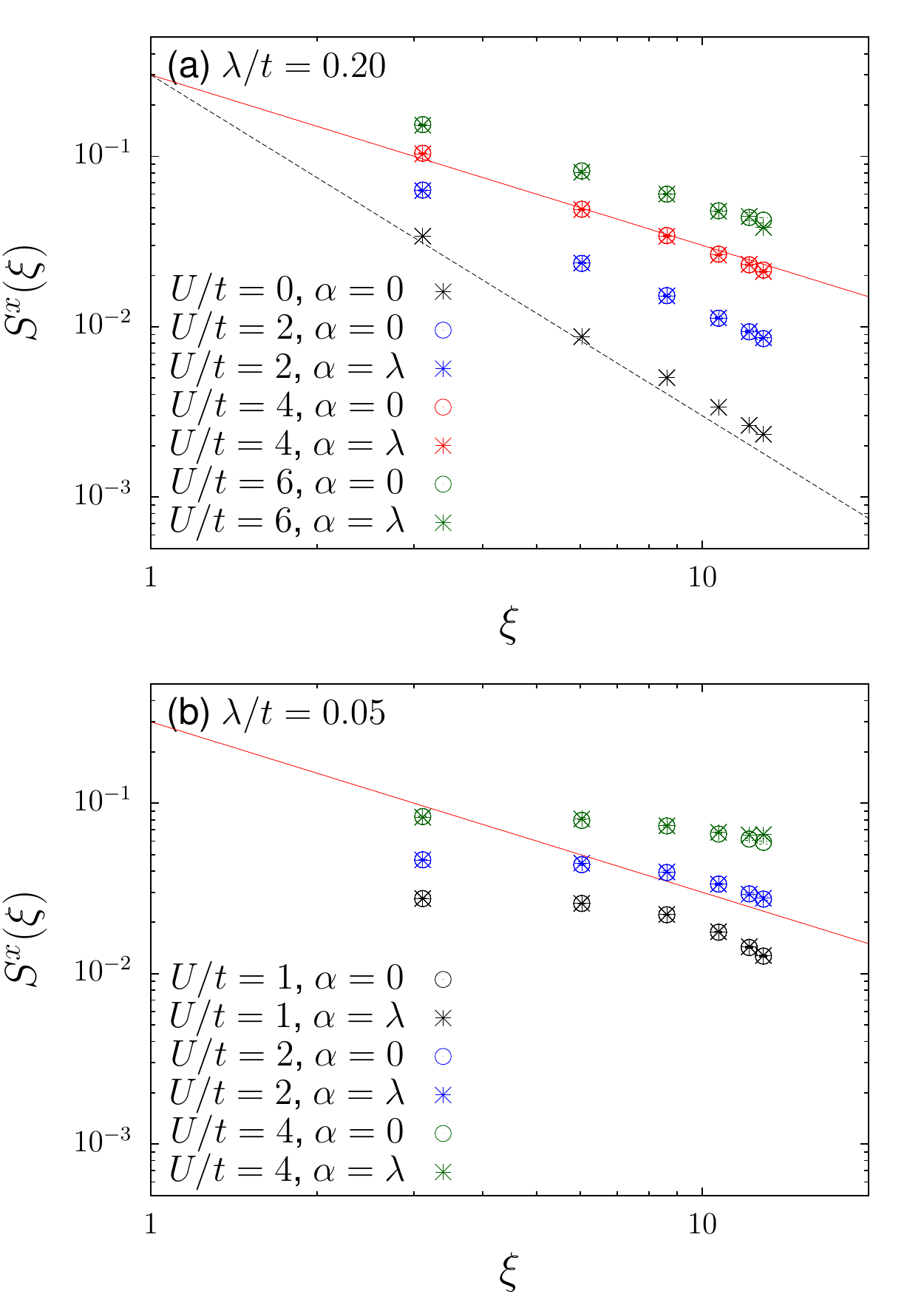}
  \caption{\label{fig:spincorrelationsU} (Color online) Projective CT-INT
    results for the transverse spin correlations for different interactions
    $U$. Here, $\lambda/t=0.2$, $L=13$, $L_\perp=64$,  $\theta t = 30$. 
    The dashed (solid) lines illustrate the $1/r^2$ ($1/r$) decay expected for $K=1$
    ($K=1/2$).}
\end{figure}

For odd values of the ribbon length $L$, we indeed find a well-defined
power-law decay at large $\xi$ with an exponent very close to 2 both for
$L=65$ and $L=125$. The expected $1/r^2$ power law is also observed for
longitudinal spin correlations (not shown) and charge correlations
[Fig.~\ref{fig:correlationsU}(b)]. In contrast, for even values of $L$, the
correlations decay exponentially beyond a certain distance. This odd/even
effect has previously been reported in
Ref.~\cite{PhysRevB.83.205122}. Importantly, the helical edge states have a
symmetry-protected crossing at $k=\pi$ for even values of $L$ (see
Fig.~\ref{fig:spectraU0}). In the following, we use odd $L$ for the
calculation of correlation functions, and even $L$ for spectral functions (so
that $k=\pi$ is one of the allowed wavevectors).

\begin{figure*}[t]
  \includegraphics[width=0.9\textwidth]{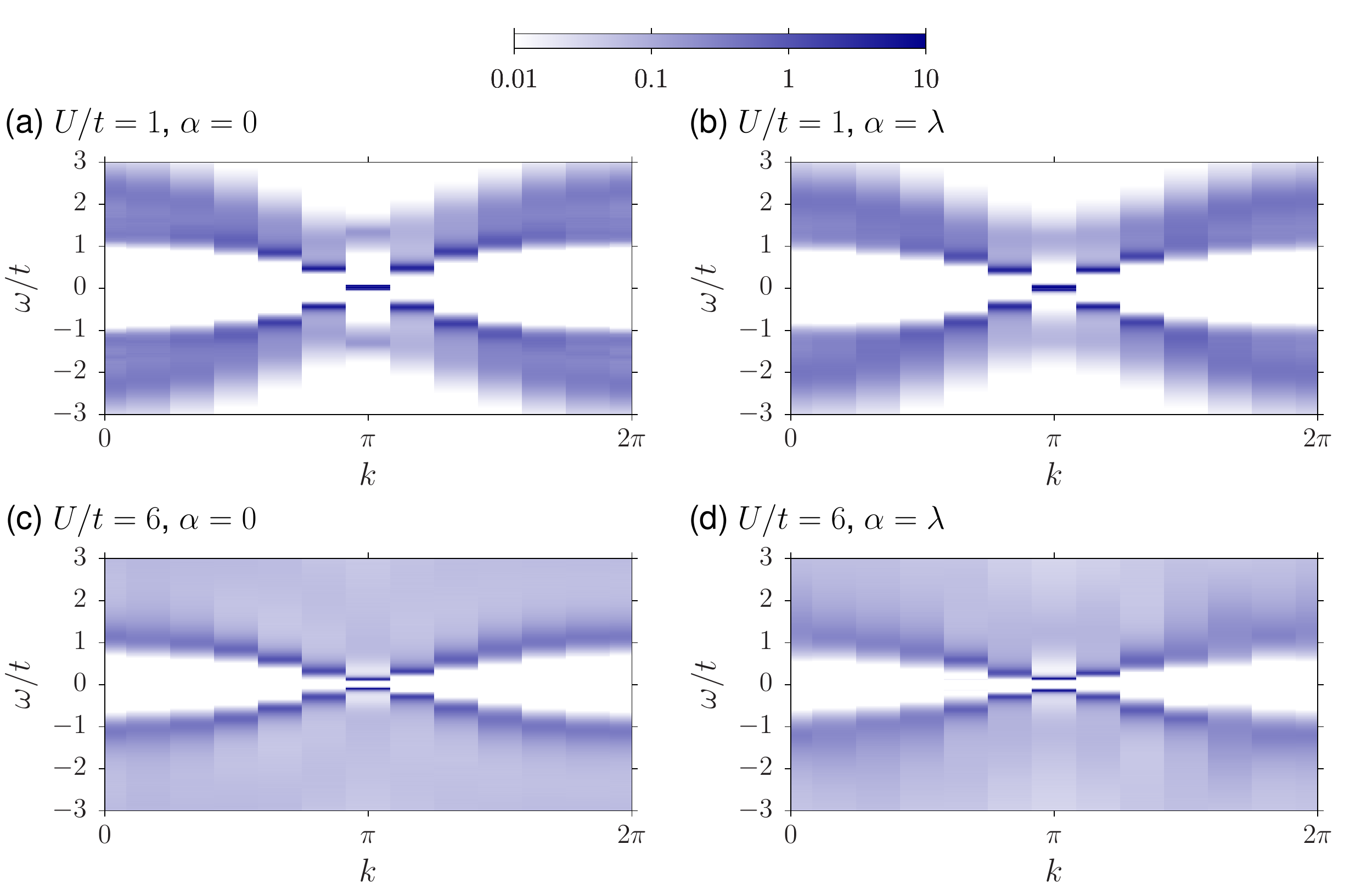}
  \caption{\label{fig:sp0.20} (Color online) CT-INT results for the
    single-particle spectral function. Here, $\lambda/t=0.2$, $L=12$,
    $L_\perp=64$, and $\beta t = 60$.}
\end{figure*}

\begin{figure*}[t]
  \includegraphics[width=0.9\textwidth]{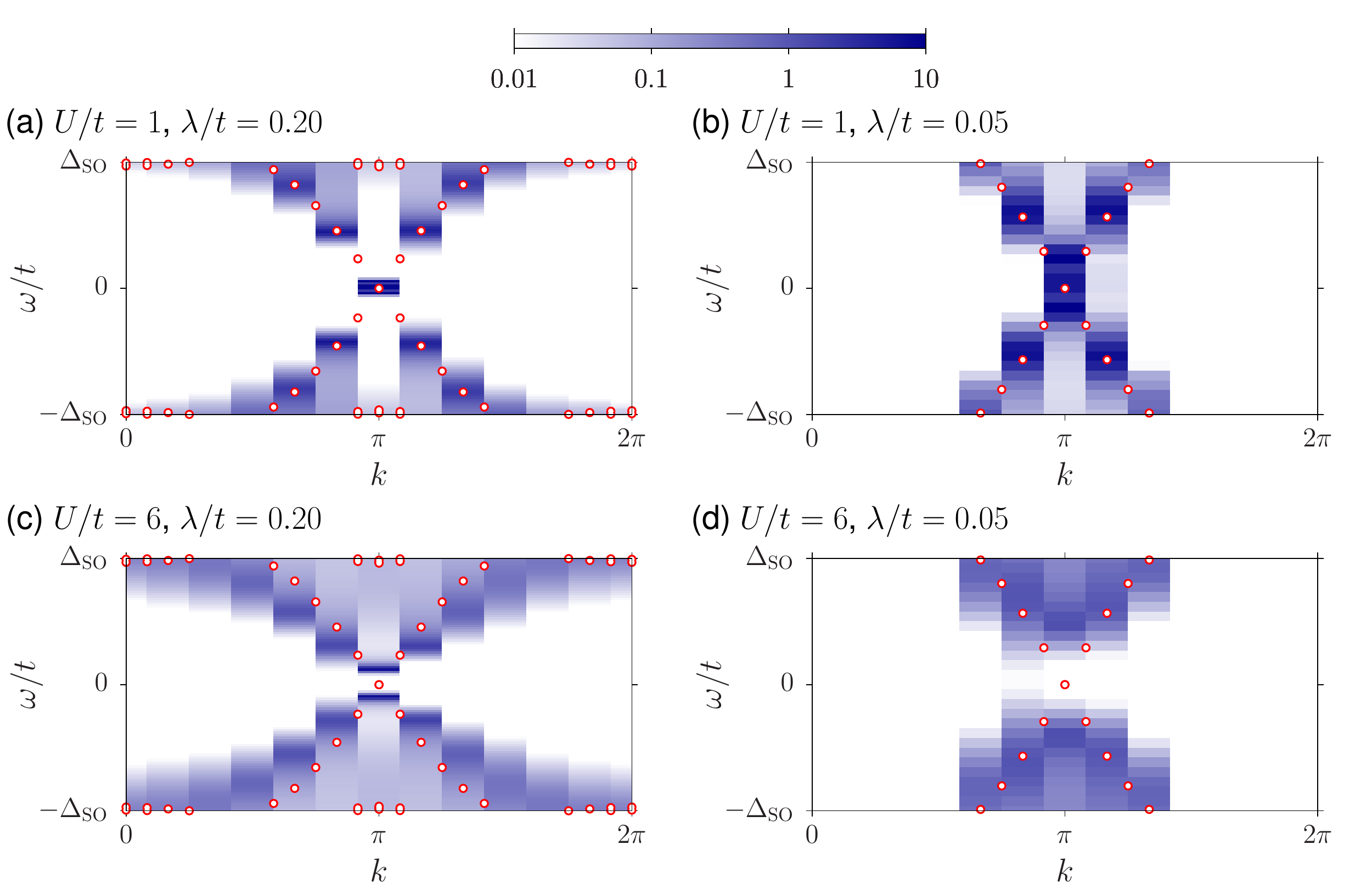}
  \caption{\label{fig:spzoom} (Color online) CT-INT results for the
    single-particle spectral function in the energy range
    $[-\Delta_\text{SO},\Delta_\text{SO}]$, where
    $\Delta_\text{SO}=3\sqrt{3}\lambda$ is the band gap of the noninteracting
    model.  Here, $\alpha=0$, $L=12$, $L_\perp=64$, and $\beta t = 60$. Symbols
    correspond to the eigenvalues of the noninteracting model with $L=24$ and
    $\alpha=0$ (see also Fig.~\ref{fig:spectraU0}).}
\end{figure*}

\subsection{Correlation functions}

Having established that we are able to see the power-law correlations
theoretically predicted for $U=0$, we consider the impact of a Hubbard
interaction on the real-space spin and charge correlations. The results
obtained with the projective CT-INT method are converged within the error
bars (if larger than the symbols) or the symbol size, respectively.

We first consider weak and moderately strong interactions on a $65\times 64$
ribbon that is large enough to reveal the asymptotic
behavior. Figure~\ref{fig:correlationsU}(a) shows the transverse spin
correlator $S^{x}(\xi)$ for different values of $U/t$. For $U=0$, we have the
same $1/r^2$ decay as in Fig.~\ref{fig:correlationsU0}. With increasing $U/t$,
the transverse spin correlations decay more slowly. Although deviations are
visible for small distances, the behavior for $\xi\gtrsim10$ fits very well
to a power-law decay $r^{-\gamma}$ with $\gamma<2$. The reduction of the
exponent as a result of repulsive interactions matches the bosonization
result of Eq.~(\ref{eq:spincorrelator}), where $\gamma=2K$ and $K$ reduces
from $K=1$ for $U=0$ to $K<1$ for $U>0$. By fitting the decay of $S^{x}(\xi)$
to the form $a/\xi^\gamma$ at large distances [there are no logarithmic
corrections to Eq.~(\ref{eq:spincorrelator})], we obtain the estimates
$K=0.882(1)$ for $U/t=1$ and $K=0.774(2)$ for $U/t=2$.  The transverse spin
correlations are very similar for $\alpha=0$ and
$\alpha=\lambda$. This result is consistent with theoretical
predictions: At the Luttinger liquid fixed point, the Rashba coupling scales
to zero.

The longitudinal spin correlations and the charge correlations are expected
to decay with the same exponent $\gamma=2$ for any value of $U$ in the
Luttinger liquid phase \cite{Ho.As.11}. Hence, we have for the charge
correlator
\begin{equation}\label{eq:chargecorrelator}
  N(r)= \las n(r) n(0) \ras 
  \sim 
  r^{-2}
  \,.
\end{equation}
Our numerical results for $N(\xi)$ in Fig.~\ref{fig:correlationsU}(b) confirm
that the exponent remains unchanged for $U>0$; the fits to a power-law give
$K=1$ within error bars. The excellent agreement of our results with
theoretical predictions illustrates that we can indeed reach the low-energy
regime, and that the bosonization results hold even for $U>\Delta_\text{SO}$.

Because the numerical effort for simulations with the CT-INT method scales
roughly as $U^3$ \cite{Rubtsov05}, we are forced to consider smaller $L$
in the strong-coupling regime ($L_\perp$ remains unchanged). Accordingly,
Fig.~\ref{fig:spincorrelationsU} shows the transverse spin correlations for
$L=13$, a system size small enough to reliably calculate the single-particle
spectral function below. The exponent for the decay of $S^{x}(\xi)$ decreases
further as we consider larger values of $U/t$. For $U/t=4$ and $U/t=6$ it
appears to be smaller than the critical value $\gamma=1$ (or $K=1/2$,
illustrated by the solid line) expected at the Mott transition. (For
$\alpha>0$, $T=0$, and half-filling, the smallest possible exponent for
power-law correlations is $\gamma=1$; beyond that, long-range order sets
in. In contrast, for $\alpha=0$, the lower bound is $\gamma=0$.)  The spin
correlations are significantly enhanced at a given $U$ for a smaller
$\lambda/t=0.05$, as shown in Fig.~\ref{fig:spincorrelationsU}(b). As pointed
out before, correlation effects on the edge states are determined by the
ratio $U/\lambda$ \cite{Ho.As.11}.

Figure~\ref{fig:correlationsU}(a) also includes results for a nonzero Rashba
coupling $\alpha=\lambda$, which are essentially identical to those for
$\alpha=0$. This suggests that the parameter $K$ is also the same.  Clearly,
for the system sizes accessible in the strong-coupling regime, we cannot
distinguish a slow power-law decay from potential long-range order.  However,
our results reveal the important fact that there is very little difference
between results obtained with and without Rashba coupling.

\subsection{Single-particle spectral function}

A key signature of the Mott transition predicted for helical liquids with
strong umklapp scattering is the opening of a gap in the single-particle
spectrum. As usual for Kosterlitz-Thouless transitions, the gap opens
exponentially as a function of $U$ close to the critical point. Moreover, in
one dimension, long-range order is only possible at zero temperature.

Here, we calculate the single-particle Green function and the corresponding
spectral function at different temperatures. We consider the spin-diagonal
and spin-averaged Green function
\begin{equation}
  G(k,\tau)
  =  
  \frac{1}{2}\la c^\dag_{k\UP}(\tau)  c^\nag_{k\UP}(0) + c^\dag_{k\DO}(\tau)  c^\nag_{k\DO}(0) \ra
\end{equation}
from which we obtain the single-particle spectral function
\begin{align}\label{eqn:akw}
  A(k,\omega)&=-\frac{1}{\pi}\mathrm{Im}\, G(k,\omega)
\end{align}
by inverting the relation (we set $\mu=0$)
\begin{equation}\label{eq:maxent}
  G(k,\tau) = \int_{-\infty}^\infty \rmd \omega
  \frac{e^{-\omega \tau}}{1+e^{-\beta\omega}} A(k,\omega)\,.
\end{equation}
with the stochastic maximum entropy method \cite{Beach04a}.

Figure~\ref{fig:sp0.20} shows results for $\lambda/t=0.2$ and $\beta t=60$.
We compare the cases of weak and strong interactions, as well as zero and
nonzero Rashba coupling. For $U/t=1$, we see almost identical spectra for
$\alpha=0$ [Fig.~\ref{fig:sp0.20}(a)] and $\alpha=\lambda$
[Fig.~\ref{fig:sp0.20}(b)]. The helical edge states are well visible inside
the spin-orbit gap, with a large spectral weight at the Fermi level for
$k=\pi$.  In the strongly correlated regime, $U/t=6$, we find  a
small gap at the Fermi level both for $\alpha=0$ [Fig.~\ref{fig:sp0.20}(c)] and
$\alpha=\lambda$ [Fig.~\ref{fig:sp0.20}(d)].

Figure~\ref{fig:spzoom} compares $A(k,\omega)$ for $\lambda/t=0.2$ and
$\lambda/t=0.05$ in the energy range set by the respective spin-orbit gap
$\Delta_\text{SO}$; the Rashba coupling has been set to zero. To reveal the
interaction effects more clearly, we also include the noninteracting
eigenvalue spectrum (points). Figure~\ref{fig:spzoom}(a) shows that for
$\lambda/t=0.2$ and $U/t=1$, the edge states resemble quite closely those of
the noninteracting system. In contrast, for $\lambda/t=0.05$ and $U/t=1$
(corresponding to roughly a four times larger effective interaction),
correlations give rise to a pseudogap at $k=\pi$. This effect is much
stronger for $U/t=6$, as shown in Figs.~\ref{fig:spzoom}(c) and~\ref{fig:spzoom}(d). Again,
the gap/pseudogap is larger (in units of $\Delta_\text{SO}$) for the more
strongly correlated case of $\lambda/t=0.05$.

Because the results for the spectral function were obtained on rather small
systems ($L=12$), it is important to investigate finite-size effects. To
avoid problems related to the analytic continuation of data with larger
statistical errors, we use an alternative way to extract the spectral
function at $\omega=0$ (see also Ref.~\cite{PhysRevB.59.4364}). Setting
$\tau=\beta/2$, we obtain from Eq.~(\ref{eq:maxent}) the relation
\begin{equation}
  \frac{1}{2}
  \lim_{\beta\to\infty}
  \beta G(k,\beta/2) = A(k,0)\,.
\end{equation}
In a finite metallic system with a $\delta$ peak at $k=\kF$ and $\om=0$,
$A(\kF,0)$ and hence $\beta G(\kF,\beta/2)$ diverges in the limit $T\to
0$. In contrast, in a gapped system, $\beta G(\kF,\beta/2)\to 0$ as $T\to0$.
 
\begin{figure}[t]
  \includegraphics[width=0.45\textwidth]{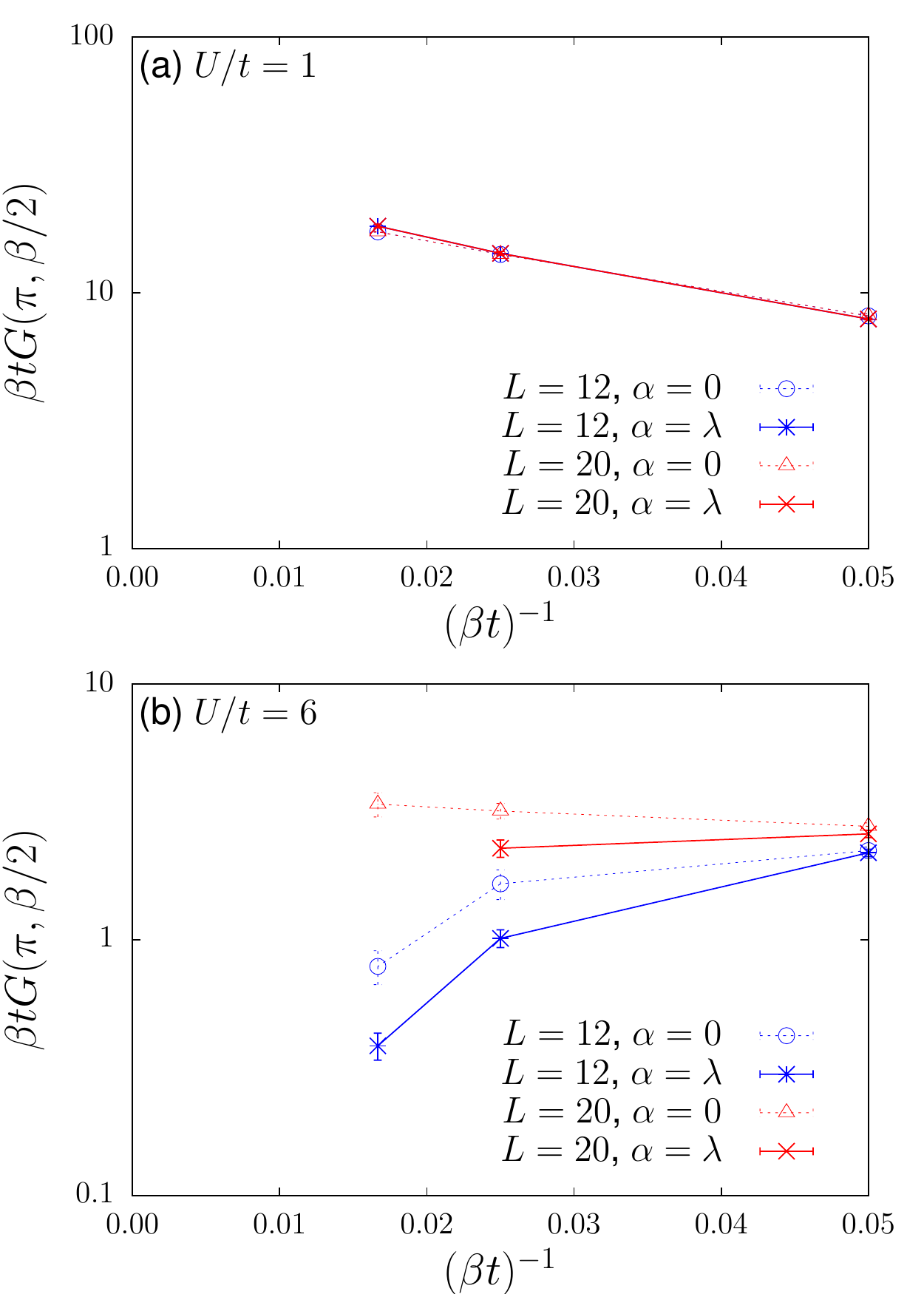}
  \caption{\label{fig:greenbeta0.20} (Color online) CT-INT results for the
    rescaled single-particle Green function at $k_\text{F}=\pi$, $\tau=\beta/2$.
    Here, $\lambda/t=0.2$, $L=12$, and $L_\perp=64$.}
\end{figure}

Figure~\ref{fig:greenbeta0.20}(a) shows $\beta G(\kF,\beta/2)$ (here
$\kF=\pi$) as a function of $1/\beta$ for $U/t=1$. For both $L=12$ and
$L=20$, $A(\kF,0)$ increases with decreasing temperature, suggesting the
absence of a gap. In contrast, for $U/t=6$ in
Fig.~\ref{fig:greenbeta0.20}(b), the results for $L=12$ reveal a gap that is
larger for $\alpha=\lambda$ than for $\alpha=0$, whereas results obtained
with and without Rashba coupling are almost identical for $U/t=1$. These
results agree with Fig.~\ref{fig:sp0.20}. Importantly, for $L=20$,
Fig.~\ref{fig:greenbeta0.20} shows a much larger spectral weight at low
temperatures; the data for $\alpha=0$ even increase with decreasing
temperature. This finding suggests that the gap seen in Fig.~\ref{fig:sp0.20}
for strong interactions and $L=12$ is a finite-size effect.  The absence of a
gap for weak interactions in Fig.~\ref{fig:spzoom}(a) (and for the
noninteracting case, not shown) implies that the gap visible in, for example,
Fig.~\ref{fig:spzoom}(c) is induced by correlations. A more detailed
discussion will be given in Sec.~\ref{sec:discussion}.

\section{Discussion}\label{sec:discussion}

{\it Without Rashba coupling}, or away from half-filling, the bosonization
results predict that all correlation functions decay with a power law at zero
temperature. With increasing $U$, transverse spin correlations decay more
slowly (the exponent is $2K$, with $K<1$ for $U>0$), but there is no
long-range order. The edge states hence remain gapless, in accordance with
the unchanged $1/r^2$ decay of charge correlations. Neglecting the
high-energy bulk states, these results hold even for very large $U$ (compared
to the bulk band gap $\Delta_\text{SO}$). Importantly, whereas a mean-field treatment of
interactions at the edge gives a magnetic state for any $U>0$,
order-parameter fluctuations destroy the long-range order and give rise to power-law
correlations even at $T=0$ \cite{PhysRevLett.107.166806}.

For a helical liquid {\it with Rashba coupling}, the same physics is expected
as long as umklapp scattering is not relevant (\ie, for $K>1/2$ and/or away
from half-filling). For $K<1/2$ at half-filling, umklapp scattering becomes a
relevant perturbation, and drives the system away from the metallic Luttinger
liquid fixed point \cite{Wu06}. The system develops long-range magnetic order
in the transverse spin direction at $T=0$, corresponding to a spontaneous
symmetry breaking with an Ising ($Z_2$) order parameter. The helical edge
states hence undergo a Mott transition and acquire a gap at the Fermi
level. The latter is allowed because the magnetic order breaks the protecting
time-reversal symmetry at the edge. In the Mott phase, density fluctuations
are exponentially suppressed. At $T>0$, the long-range order is destroyed by
thermally induced fluctuations of the order parameter, and time-reversal
symmetry is restored. We can think of the ordered state in terms of a scalar field.
The order parameter correlation length
(the average distance between thermally induced domain walls) decreases
exponentially with increasing temperature. Similar to solitons in the
Su-Schrieffer-Heeger model \cite{Su80}, domain walls (kinks) correspond
to midgap states, and the single-particle gap is 
filled in at $T>0$.

Let us compare these theoretical predictions with the numerical results
presented in Sec.~\ref{sec:results}. Figure~\ref{fig:correlationsU0}
demonstrates that, for not too large values of $U$, our simulations capture
the low-energy and long-wavelength limit where the bosonization results of
Eqs.~(\ref{eq:spincorrelator}) and~(\ref{eq:chargecorrelator}) hold. A
power-law decay persists even when $U$ becomes larger than the spin-orbit
gap, although deviations can be observed when $U\gg\Delta_\text{SO}$. For
$\lambda/t=0.2$ and $U/t\leq2$, corresponding to the Luttinger liquid regime
with $K>1/2$, the results are almost unaffected by Rashba coupling, which
is an irrelevant perturbation at the Luttinger liquid fixed point.
Transverse spin correlations are strongly enhanced with increasing $U$,
with $U/\lambda$ being the relevant parameter ratio for the low-energy
physics.

For large $U/\lambda$, spin correlations decay increasingly slowly both with
or without Rashba coupling. Although theory predicts a gapless Luttinger liquid
even at $T=0$, we observe a gap in the single-particle spectrum at low but
finite temperatures. This gap decreases with increasing system size, and is
expected to close in the thermodynamic limit. However, the accessible system
sizes are insufficient to capture the quantum fluctuations of the order
parameter that destroy the long-range order induced by the mean-field
instability \cite{PhysRevLett.107.166806}. By this mechanism, we observe a
single-particle gap even without a Rashba term that becomes larger (for a
given system size) with increasing $U/\lambda$. Similarly, a gap is observed
with Rashba coupling at $T>0$ for small system sizes. While spin correlations
are very similar with and without Rashba coupling, the spectral weight at the
Fermi level is more strongly suppressed in the presence of Rashba coupling.

The generic strong spin correlations and mean-field instability of helical
liquids---both of which are not specific to half-filling---make it very
difficult to observe clear signatures of the umklapp-induced Mott
transition numerically. In addition to the quasi-long-range order of spins,
it has been previously shown that the spectral weight at the Fermi level is
suppressed by spin-flip scattering involving bulk states
\cite{Ho.As.11,PhysRevB.83.205122}, giving rise to a
pseudogap. Distinguishing quasi-long-range order [$K\ll1$ in
Eq.~(\ref{eq:spincorrelator})] and a pseudogap for $\alpha=0$ from long-range
order and a gap for $\alpha>0$ is very difficult. 

The Mott transition in helical liquids is expected to become visible in
simulations on much larger two-dimensional lattices and at zero temperature,
for which suitable methods will have to be developed.  A possible route
toward this goal is tensor networks, which have
recently been applied to quantum Hall systems
\cite{PhysRevLett.106.156401,HoAsreview2013}. Other promising ways to detect
the Mott transition numerically include the calculation of the edge topological
invariant \cite{PhysRevB.83.085426} or the entanglement entropy
\cite{PhysRevLett.90.227902}, both of which are qualitatively (rather than
quantitatively) different in the ordered and disordered phases, respectively.

\section{Conclusions}\label{sec:conclusions}
\vspace*{-0.5em}
We studied the role of Rashba coupling for helical edge states of quantum
spin Hall insulators by simulating a Kane-Mele model with a Hubbard
interaction at the edge using quantum Monte Carlo methods. We were able to
reveal the theoretically predicted asymptotic behavior of spin and charge
correlation functions. Because of limitations in system size, we observed an
interaction-induced single-particle gap at strong interactions, which is
expected to close or be filled in in the thermodynamic limit at any finite
temperature. Quite surprisingly, our numerical data are very similar for
models with and without Rashba coupling, although the latter allows for a
Mott state with long-range magnetic order at zero temperature that breaks
time-reversal symmetry.

Our observation of a gap for large $U$ can be attributed to 
the fact that in the regime where a Mott state can exist ($K<1/2$),
spin correlations decay slower than $1/r$. On finite lattices, such a slow
decay is difficult to distinguish from mean-field order. Put differently, the
order parameter fluctuations that destroy the magnetic order for $K>1/2$ and
$T=0$ or for $K<1/2$ and $T>0$ involve length scales that are not fully
captured by the accessible system sizes. Hence, whereas numerical work has
successfully linked Mott transitions in other one- and two-dimensional
models to theoretical predictions, the Mott transition of a helical Luttinger
liquid appears to be out of reach for existing numerical methods.

While the requirement of half-filling at first
suggests that the Mott transition is not particularly relevant for experiments, our results
reveal that correlated helical edge states exhibit many of the
characteristics of finite-temperature Mott states (pronounced spin
correlations, suppressed spectral weight at the Fermi level)
even when a Mott transition is not allowed.

\vspace*{-1.5em}
{\begin{acknowledgments}%
    We are grateful to the J\"ulich Supercomputing Centre for computer time,
    and acknowledge financial support from the DFG Grants Nos.~AS120/9-1 and
    Ho 4489/3-1 (FOR 1807). We further thank L. Fritz and S. Rachel for
    helpful discussions during the SPORE13 workshop.
  \end{acknowledgments}}

\end{document}